\documentclass[12pt,subeqn]{article}

\newcommand{\EQ}{\begin{equation}}
\newcommand{\EN}{\end{equation}}

\newcommand{\ket}[1]{\left|#1\right\rangle}      

\newcommand{\bear}{\begin{eqnarray}}
\newcommand{\ear}{\end{eqnarray}}
\newcommand{\bt} { \begin{tabular} }
\newcommand{\et}{ \end{tabular} }
\newcommand{\bc} { \begin{center} }
\newcommand{\ec}{ \end{center} }

\newcommand{\btb} { \begin{table} }
\newcommand{\etb}{ \end{table} }

\begin{document}

\topmargin 0pt
\oddsidemargin 5mm
\newcommand{\NP}[1]{Nucl.\ Phys.\ {\bf #1}}
\newcommand{\PL}[1]{Phys.\ Lett.\ {\bf #1}}
\newcommand{\NC}[1]{Nuovo Cimento {\bf #1}}
\newcommand{\CMP}[1]{Comm.\ Math.\ Phys.\ {\bf #1}}
\newcommand{\PR}[1]{Phys.\ Rev.\ {\bf #1}}
\newcommand{\PRL}[1]{Phys.\ Rev.\ Lett.\ {\bf #1}}
\newcommand{\MPL}[1]{Mod.\ Phys.\ Lett.\ {\bf #1}}
\newcommand{\JETP}[1]{Sov.\ Phys.\ JETP {\bf #1}}
\newcommand{\TMP}[1]{Teor.\ Mat.\ Fiz.\ {\bf #1}}

\renewcommand{\thefootnote}{\fnsymbol{footnote}}

\newpage
\setcounter{page}{0}
\begin{titlepage}
\begin{flushright}
UFSCARF-TH-07-03
\end{flushright}
\vspace{0.5cm}
\begin{center}
{\large Solution of the $SU(N)$ Vertex Model with Non-Diagonal Open Boundaries}\\
\vspace{1cm}
{\large W. Galleas and M.J. Martins} \\
\vspace{1cm}
{\em Universidade Federal de S\~ao Carlos\\
Departamento de F\'{\i}sica \\
C.P. 676, 13565-905~~S\~ao Carlos(SP), Brasil}\\
\end{center}
\vspace{0.5cm}

\begin{abstract}
We diagonalize the double-row transfer matrix of the $SU(N)$ 
vertex model for certain classes of non-diagonal
boundary conditions. We derive explicit expressions for the 
corresponding eigenvectors and eigenvalues by means
of the algebraic Bethe ansatz approach.
\end{abstract}

\vspace{.15cm}
\centerline{PACS numbers:  05.50+q, 02.30.IK}
\vspace{.1cm}
\centerline{Keywords: Algebraic Bethe Ansatz, Lattice Models, Open Boundary Conditions}
\vspace{.15cm}
\centerline{July 2004}

\end{titlepage}


\pagestyle{empty}

\newpage

\pagestyle{plain}
\pagenumbering{arabic}

\renewcommand{\thefootnote}{\arabic{footnote}}

The study of integrable models of statistical mechanics with arbitrary boundary conditions has gained 
a tremendous impulse after the work by Sklyanin \cite{SK}. This author has been able to generalize the 
quantum inverse scattering method \cite{QI} to include the important case of open boundaries. It turns out
that an exactly solved spin chain with open boundary condition can be obtained through the following
double-row transfer matrix
\EQ
T(\lambda)=\mbox{Tr}_{\cal A}\left[ K^{(+)}_{\cal A} (\lambda) {\cal T}_{\cal A}(\lambda)
K^{(-)}_{\cal A} (\lambda) {\cal T}_{\cal A}^{-1}(-\lambda) 
\right]
\label{tran}
\EN
where ${\cal T}_{\cal A}(\lambda)= {\cal L}_{{\cal A} L}(\lambda){\cal L}_{{\cal A} L-1}(\lambda)
\dots {\cal L}_{{\cal A} 1}(\lambda)$ is the standard monodromy matrix that generates the corresponding closed 
spin chain with $L$ sites \cite{QI}.

We recall that the symbol $\cal A$ denotes a $N$-dimensional auxiliary space and $\lambda$ parameterizes the
integrable manifold. The operator
${\cal L}_{{\cal A} j}(\lambda)$ represents the bulk weights of the corresponding vertex model whose
transfer matrix commutes with the spin chain Hamiltonian. The $N \times N$ matrices
$K_{\cal A}^{(\pm)}(\lambda)$ describe the interactions at the right and left ends of the open chain.
One of the simplest integrable system is the fundamental $SU(N)$ vertex model \cite{RE} whose
Boltzmann weights are given by
\EQ
{\cal L}_{{\cal A} j}(\lambda)=a(\lambda)\sum_{\alpha=1}^{N} \hat{e}_{\alpha\alpha}^{(\cal A)} \otimes \hat{e}_{\alpha\alpha}^{(j)}
+ b(\lambda)\sum_{\stackrel{\alpha,\beta=1}{\alpha \neq \beta}}^{N} \hat{e}_{\alpha\alpha}^{(\cal A)} \otimes \hat{e}_{\beta\beta}^{(j)}
+ \sum_{\stackrel{\alpha,\beta=1}{\alpha \neq \beta}}^{N} \hat{e}_{\alpha\beta}^{(\cal A)} \otimes \hat{e}_{\beta\alpha}^{(j)} 
\label{loper}
\EN
where $a(\lambda)=\lambda+1$, $b(\lambda)=\lambda$ and $\hat{e}_{ij}^{(V)}$ are the usual Weyl matrices 
acting on the space $V$. A quite general class of open boundary conditions for this vertex model is  represented
by the following $K$-matrices \cite{DE,KU,NEP}
\EQ
K_{\cal A}^{(\pm)}(\lambda)= M_{\cal A}^{(\pm)} D_{\cal A}^{(\pm)} (\lambda) 
\left [ M_{\cal A}^{(\pm)} \right ]^{-1}~~~~D_{\cal A}^{(\pm)}(\lambda)= 
\sum_{\alpha=1}^{N} \varepsilon_{\alpha}^{(\pm)}(\lambda) \hat{e}_{\alpha \alpha}^{(\cal A)}
\label{kma}
\EN

The elements of the $N \times N$ matrices $M_{\cal A}^{(\pm)}$ are arbitrary c-numbers and the functions
$\varepsilon_{\alpha}^{(\pm)}(\lambda)$ are given by

\begin{equation}
\label{par}
\varepsilon_{\alpha}^{(-)} (\lambda)=\cases{
\xi_{-}+\lambda \;\;\;\; \alpha=1,\dots ,p \cr
\xi_{-}-\lambda \;\;\;\; \alpha=p+1,\dots ,N \cr}
\;\;\;\;\;
\varepsilon_{\alpha}^{(+)} (\lambda)=\cases{
\xi_{+}-\frac{N}{2}-\lambda \;\;\;\; \alpha=1,\dots ,p \cr
\xi_{+}+\frac{N}{2}+\lambda \;\;\;\; \alpha=p+1,\dots ,N \cr}
\end{equation}
where $p$ is an integer with values on the interval $1 \leq p \leq N$
 and $\xi_{\pm}$ are free-parameters. Here we emphasize
that each $K$-matrix (\ref{kma},\ref{par}) has altogether $2N-1$ arbitrary parameters characterizing the interactions at the
appropriate boundary.

The diagonalization of the transfer matrix (\ref{tran},\ref{loper}) for general non-diagonal
$K$-matrices (\ref{kma},\ref{par}) is a tantalizing problem due to the difficulty of
finding a suitable reference state to perform a Bethe ansatz analysis. However, progress on this matter
has recently been done in the literature, most of it concentrated on the eight \cite{CH} and six \cite{NEP,CA,JI}
vertex models. The $Z_N$ Belavin model is to our knowledge the only multistate vertex system investigated so far
with non-diagonal open boundaries \cite{JA}.  Though its bulk weights are known to reduce in the isotropic
limit to those of the $SU(N)$ vertex model (\ref{loper}), the same does not occur for the boundary $K$-matrices.
In fact, the elliptic $K$-matrices associated to the $Z_N$ Belavin model \cite{CH1} have
fewer free-parameters, which totals $N+1$, as compared to that contained
in the isotropic $K$-matrices (\ref{kma},\ref{par}). Therefore, for $N \geq 3$ the $SU(N)$ vertex model
with open boundaries is indeed a genuine integrable system that deserves to be studied independently. We suspect that
this situation extends to many isotropic integrable vertex models based on higher rank symmetries.

The purpose of this work is to show that the diagonalization of the transfer matrix (\ref{tran},\ref{loper})
of the $SU(N)$ vertex model in the case $M_{\cal A}^{(+)}= M_{\cal A}^{(-)}$ can be mapped on a similar
eigenvalue problem with the diagonal boundaries $D_{\cal A}^{(\pm)}(\lambda)$.  This constraint
does not imply that the right and left $K$-matrices  are the same because the parameters $\xi_{\pm}$ are
still unrelated.  This observation not only allows us to solve the eigenvalue problem for
$2N$ independent boundary parameters but also makes 
it possible the relation between eigenvectors of
seemly different open boundaries. In order to see that we insert the terms 
$M_{\cal A}^{(-)} 
\left [ M_{\cal A}^{(-)} \right ]^{-1}$ all over the 
trace of the double-row transfer matrix (\ref{tran}), permitting us to rewrite it as
\EQ
T(\lambda)=\mbox{Tr}_{\cal A} \left[ D^{(+)}_{\cal A} (\lambda) \widetilde{{\cal T}}_{\cal A}(\lambda)
D^{(-)}_{\cal A} (\lambda) \widetilde{{\cal T}}_{\cal A}^{-1}(-\lambda) \right ]
\label{tran1}
\EN
where the new monodromy $\widetilde{{\cal T}}_{\cal A}(\lambda)= \widetilde{{\cal L}}_{{\cal A} L}(\lambda)
\widetilde{{\cal L}}_{{\cal A} L-1}(\lambda)
\dots \widetilde{{\cal L}}_{{\cal A} 1}(\lambda)$
whose gauge transformed $\widetilde{\cal L}$-operators are given by
\EQ
\widetilde{{\cal L}}_{{\cal A} j}=\left [ M_{\cal A}^{(-)} \right ]^{-1} {\cal L}_{{\cal A} j}(\lambda) 
M_{\cal A}^{(-)}
\label{tra1}
\EN

Further progress is made by reversing the gauge transformation (\ref{tra1}) with the help of the
following quantum space transformation
\EQ
U_j^{-1} 
\widetilde{{\cal L}}_{{\cal A} j}(\lambda) U_j= 
{\cal L}_{{\cal A} j}(\lambda)
\label{tra2}
\EN
where $U_j= \mbox{Id} \otimes \dots \mbox{Id} \otimes \underbrace{M_{A}^{(-)}}_{j-\mbox{th}} \otimes \mbox{Id} \dots \otimes \mbox{Id}$ and
$\mbox{Id}$ is the  $ N \times N$ identity matrix. This remarkable
property \cite{RM} can now be used to define a new double-row operator $\bar{T}(\lambda)$
\EQ
\bar{T}(\lambda)= \displaystyle {\prod_{j=1}^{L}} U_j^{-1} \; T(\lambda) \;
\displaystyle {\prod_{j=1}^{L} U_j }
\label{cano}
\EN

By using the canonical transformation (\ref{cano}) together with the assumed constraint
$M_{\cal A}^{(+)}=M_{\cal A}^{(-)}$ between the right and the left $K$-matrices we find that
$\bar{T}(\lambda)$ becomes
\EQ
\bar{T}(\lambda)=\mbox{Tr}_{\cal A}\left[ D^{(+)}_{\cal A} (\lambda) {\cal T}_{\cal A}(\lambda)
D^{(-)}_{\cal A} (\lambda) {\cal T}_{\cal A}^{-1}(-\lambda) 
\right]
\label{tran2}
\EN
which is precisely the double-row transfer matrix of the $SU(N)$ vertex model with
diagonal $K$-matrices $D_{\cal A}^{(\pm)}(\lambda)$.

As a consequence of that the operators $T(\lambda)$ and $\bar{T}(\lambda)$ share the same eigenvalues and
furthermore if $\ket{\bar{\psi}}$ is an eigenstate of $\bar{T}(\lambda)$ then the corresponding eigenvector $\ket{\psi}$
of $T(\lambda)$ is $\displaystyle \prod_{j=1}^{L} U_j \ket{\bar{\psi}}$. From now
on our main task consists therefore in diagonalizing the double-row transfer matrix $\bar{T}(\lambda)$. In this
case the associated open boundaries $D_{\cal A}^{(\pm)}(\lambda)$ are diagonal and such eigenvalue
problem can be tackled by standard nested Bethe ansatz approach. By now this procedure has been well
explained in the literature, see for instance refs. \cite{FO,CH2}, and here we shall restrict ourselves
in presenting only the essential steps of the solution. We first note that diagonal boundaries permit us
to use as pseudovacuum the usual ferromagnetic state
\EQ
\ket{\bar{\psi}_0}= \prod_{j=1}^{L} \otimes \ket{0}_j,~~~
\ket{0}_{j}=\left(\begin{array}{c}
                1 \\
                0  \\
                \vdots \\
                0 
                \end{array}\right)_N,
\label{vec}
\EN

The next step is to write convenient commutation rules for the elements of the double transition
operator $\Upsilon_{\cal A}(\lambda)=
\widetilde{{\cal T}}_{{\cal A}}(\lambda) D_{\cal A}^{(-)}(\lambda) 
\widetilde{{\cal T}}_{{\cal A}}^{-1}(-\lambda) $ which satisfies the following quadratic relation \cite{SK}
\EQ
{\cal L}_{12}(\lambda-\mu) \Upsilon_1(\lambda)
{\cal L}_{21}(\lambda+\mu) \Upsilon_2(\mu)= \Upsilon_2(\lambda)
{\cal L}_{12}(\lambda+\mu) \Upsilon_1(\mu)
{\cal L}_{21}(\lambda-\mu) 
\label{quad}
\EN

We proceed by looking for a representation of the operator $\Upsilon_{\cal A}(\lambda)$ that is capable to
distinguish potential creation and annihilation fields over the state $\ket{0}$. Previous experience with
nested Bethe ansatz diagonalization of the $SU(N)$ vertex model \cite{RE} suggests us the form
\EQ
\Upsilon_{\cal A}(\lambda)=\left(\begin{array}{cccc}
                A(\lambda) & B_{1}(\lambda) & \cdots & B_{N-1}(\lambda) \\
                C_{1}(\lambda) & D_{11}(\lambda) & \cdots & D_{1N-1}(\lambda) \\
                \vdots & \vdots & \ddots & \vdots \\
                C_{N-1}(\lambda) & D_{N-11}(\lambda) & \cdots & D_{N-1 N-1}(\lambda) \\
                \end{array}\right)_{N \times N}.
\label{repre}
\EN

From Eqs.(\ref{quad},\ref{repre}) it follows that out of $N^4$ possible commutation rules there
exists three families that are of great use, namely
\begin{eqnarray}
\label{co1}
A(\lambda) B_{j}(\mu) &=& \frac{a(\mu - \lambda)}{b(\mu - \lambda)} \frac{b(\mu + \lambda)}{a(\mu + \lambda)}
B_{j}(\mu) A(\lambda) - \frac{b(2\mu )}{a(2\mu )} \frac{1}{b(\mu - \lambda)} B_{j}(\lambda) A(\mu) \nonumber \\
&-&\frac{1}{a(\lambda + \mu)} B_{i}(\lambda) \widetilde{D}_{ij}(\mu) \\
\label{co2}
\widetilde{D}_{ij} (\lambda) B_{k}(\mu) &=& \frac{\left[\hat{r}^{(1)} (\lambda +\mu +1) \right]^{id}_{ef} \left[\hat{r}^{(1)} (\lambda -\mu) \right]^{fg}_{kj}}{b(\lambda+\mu+1) b(\lambda - \mu)}
B_{d}(\mu) \widetilde{D}_{eg} (\lambda) -\frac{\left[\hat{r}^{(1)} (2\lambda +1) \right]^{id}_{ej}}{a(2\lambda) b(\lambda - \mu)} B_{d}(\lambda) \widetilde{D}_{ek}(\mu) \nonumber \\
&+& \frac{b(2\mu )}{a(2\mu )}  \frac{\left[\hat{r}^{(1)}  (2\lambda +1) \right]^{id}_{kj}}{a(2\lambda) a(\lambda + \mu)} B_{d}(\lambda) A(\mu) \\
\label{co3}
B_{i}(\lambda) B_{j}(\mu) &=& \frac{\left[\hat{r}^{(1)} (\lambda -\mu) \right]^{cd}_{ji}}{a(\lambda -\mu)} B_{c}(\mu) B_{d}(\lambda)
\end{eqnarray}
where $\left[\hat{r}^{(1)} (\lambda)\right]^{ij}_{kl}$ are the matrix elements of the
$SU(N-1)$ ${\mathcal L}$-operator \footnote{Here we recall that we have used the convention
$\hat{r}^{(l)}_{12}(\lambda) = \displaystyle{\sum_{abcd}^{N-l} \left[\hat{r}^{(l)} (\lambda)\right]^{ab}_{cd} \hat{e}_{ac}^{(1)} \otimes \hat{e}_{bd}^{(2)} }$.}
(\ref{loper}) and the new $\widetilde{D}_{ij} (\lambda)$  operators are conveniently defined by the relation
\begin{equation}
\widetilde{D}_{ij} (\lambda) = D_{ij} (\lambda) - \frac{\delta_{ij}}{a(2\lambda)} A(\lambda)
\end{equation}

Yet another important element is the action of the fields $A(\lambda), \widetilde{D}_{ij} (\lambda)$ and $C_{i}(\lambda)$
on the reference state $\ket{0}$. This can be computed by using the triangularity property
of ${\mathcal L}_{{\mathcal A} j } (\lambda)$ upon $\ket{0}_{j}$ and they are given by
\begin{eqnarray}
A(\lambda) \ket{0} &=& \varepsilon_{1}^{(-)}(\lambda) \left[a(\lambda)\right]^{2L} \ket{0} \nonumber \\
\widetilde{D}_{ij} (\lambda) \ket{0} &=& \left[b(\lambda) \right]^{2L} \left[ \varepsilon^{(-)}_{i+1}(\lambda) - \frac{\varepsilon^{(-)}_{1}(\lambda)}{a(2 \lambda)} \right] \delta_{ij} \ket{0} \nonumber \\
C_{i}(\lambda) \ket{0} &=& 0
\end{eqnarray}

We have now gathered the basic ingredients to perform an algebraic Bethe ansatz analysis. We suppose
that further eigenstates $\ket{\bar{\psi}_{n_1}}$ of $\bar{T}(\lambda)$ can be put into the following
structure
\EQ
\ket{\bar{\psi}_{n_1}}=B_{a_{1}}(\lambda_{1}^{(1)})\dots
{B}_{a_{n_{1}}}(\lambda_{n_1}^{(1)})
{\cal F}^{a_{n_{1}} \dots a_{1}} \ket{0}
\label{multi}
\EN
where the indices $a_j$ run over $N-1$ possible values and the rapidities $\{ \lambda_j^{(1)} \}$ will
be determined by solving the eigenvalue equation
\EQ
\left[ \varepsilon^{(+)}_{1}(\lambda) + \frac{1}{a(2 \lambda)}\sum_{\alpha=1}^{N-1} \varepsilon^{(+)}_{\alpha+1}(\lambda) \right]
A(\lambda) \ket{\bar{\psi}_{n_1}} + \sum_{\alpha=1}^{N-1} \varepsilon^{(+)}_{\alpha+1}(\lambda) \widetilde{D}_{\alpha \alpha}(\lambda)
\ket{\bar{\psi}_{n_1}} = \Lambda(\lambda) \ket{\bar{\psi}_{n_1}}
\EN

By carrying on the fields $A(\lambda)$ and $\widetilde{D}_{ii} (\lambda)$ over the multiparticle state
(\ref{multi}) we generate terms that are proportional to $\ket{\bar{\psi}_{n_1}}$ and those that
are not, denominated unwanted terms. The first ones contribute to the eigenvalue $\Lambda(\lambda)$
and are obtained by keeping only the first terms of the commutation rules (\ref{co1}-\ref{co3}) and
by requiring that the coefficients ${\cal F}^{a_{n_{1}} \dots a_{1}}$ are eigenstates of an auxiliary
double-row operator $\bar{T}^{(1)}(\lambda, \{\lambda_{j}^{(1)} \})$ given by
\begin{equation}
\bar{T}^{(1)} (\lambda, \{ \lambda_{j}^{(1)} \}) = \mbox{Tr}_{A^{(1)}} \left[ D_{{\mathcal A}^{(1)}}^{(1,+)}(\lambda) {\mathcal T}^{(1)}_{{\mathcal A}^{(1)}}
(\lambda, \{ \lambda_{j}^{(1)} \}) D_{{\mathcal A}^{(1)}}^{(1,-)}(\lambda) \widetilde{{\mathcal T}}^{(1)}_{{\mathcal A}^{(1)}}
(\lambda, \{ \lambda_{j}^{(1)} \}) \right]
\end{equation}
such that
\begin{eqnarray}
{\mathcal T}^{(1)}_{{\mathcal A}^{(1)}} (\lambda, \{ \lambda_{j}^{(1)} \}) &=& \hat{r}_{{\mathcal A}^{(1)}  1}^{(1)}
(\lambda +\lambda_{1}^{(1)}+1) \dots \hat{r}_{{\mathcal A}^{(1)}  a_{n_1}}^{(1)}
(\lambda +\lambda_{a_{n_1}}^{(1)}+1) \nonumber \\
\widetilde{{\mathcal T}}^{(1)}_{{\mathcal A}^{(1)}} (\lambda, \{ \lambda_{j}^{(1)} \}) &=&
\hat{r}_{{\mathcal A}^{(1)} a_{n_1}}^{(1)} (\lambda -\lambda_{a_{n_1}}^{(1)}) \dots
\hat{r}_{{\mathcal A}^{(1)} 1}^{(1)} (\lambda -\lambda_{1}^{(1)})
\end{eqnarray}
where ${\mathcal A}^{(1)} \in C^{N-1}$ and the associated $K$-matrices are
\begin{equation}
D_{{\mathcal A}^{(1)}}^{(1,+)}(\lambda)= \sum_{\alpha=1}^{N-1} \varepsilon_{\alpha+1}^{(+)} (\lambda) \; \hat{e}_{\alpha \alpha}^{({\mathcal A}^{(1)})}
\;\;\;\;
D_{{\mathcal A}^{(1)}}^{(1,-)}(\lambda)= \sum_{\alpha=1}^{N-1} \left[ \varepsilon_{\alpha+1}^{(-)} (\lambda)
-\frac{\varepsilon_{1}^{(-)} (\lambda)}{a(2 \lambda)} \right] \; \hat{e}_{\alpha \alpha}^{({\mathcal A}^{(1)})}
\end{equation}

The diagonalization of the inhomogeneous operator $\bar{T}^{(1)} (\lambda, \{ \lambda_{j}^{(1)} \})$ is
implemented by a second Bethe ansatz which is once again parameterized by a new set of rapidities
$\lambda_{1}^{(2)} \dots \lambda_{n_2}^{(2)}$. By keeping on going this procedure we are able to relate the
eigenvalues  $\Lambda^{(l)}(\lambda, \{ \lambda_{j}^{(l)} \})$ of the transfer matrix
$\bar{T}^{(l)} (\lambda, \{ \lambda_{j}^{(l)} \})$ at the nearest neighbor steps $l$ and $l+1$. Since the
commutation relations for the elements of the corresponding double transition operator is similar to that exhibited
in Eqs. (\ref{co1}-\ref{co3}) it is not difficult to derive the following recursive relation
\begin{eqnarray}
\label{eig}
&& \Lambda^{(l)}(\lambda, \{ \lambda_{j}^{(l)} \}) =
Q^{(l)}(\lambda) \prod_{i=1}^{n_l} a(\lambda +\lambda_{i}^{(l)}+l) a(\lambda -\lambda_{i}^{(l)})
\prod_{i=1}^{n_{l+1}} \frac{a(\lambda_{i}^{(l+1)} - \lambda)}{b(\lambda_{i}^{(l+1)} - \lambda)}
\frac{b(\lambda + \lambda_{i}^{(l+1)} +l)}{a(\lambda + \lambda_{i}^{(l+1)} +l)} \nonumber \\
&+& \prod_{i=1}^{n_l} b(\lambda +\lambda_{i}^{(l)}+l) b(\lambda -\lambda_{i}^{(l)})
\prod_{i=1}^{n_{l+1}} \frac{1}{b(\lambda -\lambda_{i}^{(l+1)})} \frac{1}{b(\lambda +\lambda_{i}^{(l+1)}+l+1)}
\Lambda^{(l+1)}(\lambda, \{ \lambda_{j}^{(l+1)} \}) \nonumber \\
\end{eqnarray}
where the functions $Q^{(l)} (\lambda)$ are given by
\begin{equation}
Q^{(l)} (\lambda)=\cases{
\frac{\lambda (\lambda +\frac{N}{2})}{(\lambda + \frac{l}{2})(\lambda + \frac{l+1}{2})} \left( \xi_{-} +\lambda \right) \left( \frac{N}{2} -p +\xi_{+} -\lambda \right)  \;\;\;\;\; l=0,\dots , p-1 \cr
\frac{\lambda (\lambda +\frac{N}{2})}{(\lambda + \frac{l}{2})(\lambda + \frac{l+1}{2})} \left( \xi_{-} -\lambda -p \right) \left( \frac{N}{2} +\xi_{+} +\lambda \right)  \;\;\;\;\; l=p, \dots , N-1 \cr}
\end{equation}

For sake of consistency with our original eigenvalue problem we set $ \lambda_{j}^{(0)} \equiv 0$ for
$j=1,\dots,n_0 =L$. By the same token the unwanted terms generated in the eigenvalue problem of the
double-row operator $\bar{T}^{(l)} (\lambda, \{ \lambda_{j}^{(l)} \})$ are cancelled out provided that
the variables $\{ \lambda_{j}^{(l+1)} \}$ satisfy the nested Bethe ansatz equations
\begin{eqnarray}
\label{bae}
&& \prod_{i=1}^{n_{l-1}} \frac{a(\lambda_{k}^{(l)} +\lambda_{i}^{(l-1)} +l-1 )}{b(\lambda_{k}^{(l)} +\lambda_{i}^{(l-1)} +l-1 )}
\frac{a(\lambda_{k}^{(l)} -\lambda_{i}^{(l-1)})}{b(\lambda_{k}^{(l)} -\lambda_{i}^{(l-1)})}
\frac{Q^{(l-1)}(\lambda_{k}^{(l)})}{Q^{(l)}(\lambda_{k}^{(l)})}
\frac{b(2 \lambda_{k}^{(l)} +l-1 )}{a(2 \lambda_{k}^{(l)} +l )} =  \nonumber \\
&& \prod_{j \neq k}^{n_{l}} -\frac{a( \lambda_{k}^{(l)} + \lambda_{j}^{(l)} +l)}{b( \lambda_{k}^{(l)} + \lambda_{j}^{(l)} +l-1)}
\frac{a( \lambda_{k}^{(l)} - \lambda_{j}^{(l)})}{a( \lambda_{j}^{(l)} - \lambda_{k}^{(l)})}
\prod_{j=1}^{n_{l+1}} \frac{a(\lambda_{j}^{(l+1)} - \lambda_{k}^{(l)})}{b(\lambda_{j}^{(l+1)} - \lambda_{k}^{(l)})}
\frac{b(\lambda_{j}^{(l+1)} + \lambda_{k}^{(l)} +l)}{a(\lambda_{j}^{(l+1)} + \lambda_{k}^{(l)} +l)} \nonumber \\
\end{eqnarray}

Explicit results are now obtained by iterating Eqs. (\ref{eig},\ref{bae}) beginning at $l=0$ until we
reach the step $l=N-2$. At such final step one has to diagonalize an
inhomogeneous six vertex model with open boundaries
by adapting previous results obtained by Sklyanin \cite{SK}. Putting together all that 
and by making the convenient displacements $\lambda_{j}^{(l)} \rightarrow \lambda_{j}^{(l)} -\frac{l}{2} $
we find that the final result for the eigenvalues $\Lambda(\lambda)$, up to a normalization
factor of value $(1-\lambda^2)^L$, are given by the expression
\begin{eqnarray}
&&\Lambda (\lambda) = Q^{(0)} (\lambda) \left[a(\lambda) \right]^{2L}
\prod_{i=1}^{n_1} \frac{(\lambda - \lambda_{i}^{(1)} -\frac{1}{2} )}{(\lambda - \lambda_{i}^{(1)} +\frac{1}{2} )} \frac{(\lambda + \lambda_{i}^{(1)} -\frac{1}{2} )}{(\lambda + \lambda_{i}^{(1)} +\frac{1}{2} )} \nonumber \\
&+&\left[b(\lambda) \right]^{2L} \sum_{l=1}^{N-2} Q^{(l)} (\lambda) \prod_{i=1}^{n_l} \frac{(\lambda - \lambda_{i}^{(l)} +\frac{l+2}{2} )}{(\lambda - \lambda_{i}^{(l)} +\frac{l}{2} )} \frac{(\lambda + \lambda_{i}^{(l)} +\frac{l+2}{2} )}{(\lambda + \lambda_{i}^{(l)} +\frac{l}{2} )}
\prod_{i=1}^{n_{l+1}} \frac{(\lambda - \lambda_{i}^{(l+1)} +\frac{l-1}{2} )}{(\lambda - \lambda_{i}^{(l+1)} +\frac{l+1}{2} )} \frac{(\lambda + \lambda_{i}^{(l+1)} +\frac{l-1}{2} )}{(\lambda + \lambda_{i}^{(l+1)} +\frac{l+1}{2} )}
\nonumber \\
&+& \left[b(\lambda) \right]^{2L} Q^{(N-1)} (\lambda) \prod_{i=1}^{n_{N-1}} \frac{(\lambda - \lambda_{i}^{(N-1)} +\frac{N+1}{2} )}{(\lambda - \lambda_{i}^{(N-1)} +\frac{N-1}{2} )}
\frac{(\lambda + \lambda_{i}^{(N-1)} +\frac{N+1}{2} )}{(\lambda + \lambda_{i}^{(N-1)} +\frac{N-1}{2} )}
\label{einre}
\end{eqnarray}

while the Bethe ansatz roots $\{ \lambda_{i}^{(1)}, \dots, \lambda_{i}^{(N-1)} \}$ satisfy the following system
of non-linear equations
\begin{eqnarray}
\left[ \frac{(\lambda_{k}^{(1)} +\frac{1}{2} )}{(\lambda_{k}^{(1)} -\frac{1}{2} )} \right]^{2L} \Theta^{(1)} (\lambda_{k}^{(1)}) &=&
\prod_{j \neq k}^{n_1} \frac{(\lambda_{k}^{(1)} - \lambda_{j}^{(1)} +1 )}{(\lambda_{k}^{(1)} - \lambda_{j}^{(1)} -1 )} \frac{(\lambda_{k}^{(1)} + \lambda_{j}^{(1)} +1 )}{(\lambda_{k}^{(1)} + \lambda_{j}^{(1)} -1 )} \nonumber \\
&\times& \prod_{j=1}^{n_{2}} \frac{( \lambda_{j}^{(2)} -\lambda_{k}^{(1)} +\frac{1}{2} )}{( \lambda_{j}^{(2)} -\lambda_{k}^{(1)} -\frac{1}{2} )} \frac{( \lambda_{j}^{(2)} +\lambda_{k}^{(1)} -\frac{1}{2} )}{( \lambda_{j}^{(2)} +\lambda_{k}^{(1)} +\frac{1}{2} )}
\label{betre1}
\end{eqnarray}
\begin{eqnarray}
\prod_{i=1}^{n_{l-1}} \frac{(\lambda_{k}^{(l)} - \lambda_{i}^{(l-1)}  +\frac{1}{2} )}{(\lambda_{k}^{(l)}- \lambda_{i}^{(l-1)} -\frac{1}{2} )} \frac{(\lambda_{k}^{(l)} + \lambda_{i}^{(l-1)}  +\frac{1}{2} )}{(\lambda_{k}^{(l)}+ \lambda_{i}^{(l-1)} -\frac{1}{2} )} \Theta^{(l)} (\lambda_{k}^{(l)}) &=&
\prod_{j \neq k}^{n_l} \frac{(\lambda_{k}^{(l)} - \lambda_{j}^{(l)} +1 )}{(\lambda_{k}^{(l)} - \lambda_{j}^{(l)} -1 )} \frac{(\lambda_{k}^{(l)} + \lambda_{j}^{(l)} +1 )}{(\lambda_{k}^{(l)} + \lambda_{j}^{(l)} -1 )} \nonumber \\
&\times& \prod_{j=1}^{n_{l+1}} \frac{( \lambda_{j}^{(l+1)} -\lambda_{k}^{(l)} +\frac{1}{2} )}{( \lambda_{j}^{(l+1)} -\lambda_{k}^{(l)} -\frac{1}{2} )} \frac{( \lambda_{j}^{(l+1)} +\lambda_{k}^{(l)} -\frac{1}{2} )}{( \lambda_{j}^{(l+1)} +\lambda_{k}^{(l)} +\frac{1}{2} )} \nonumber \\
&& \;\;\;\;\;  l=2,\dots, N-2
\label{betre3}
\end{eqnarray}
\begin{eqnarray}
\prod_{i=1}^{n_{N-2}} \frac{(\lambda_{k}^{(N-1)} - \lambda_{i}^{(N-2)}  +\frac{1}{2} )}{(\lambda_{k}^{(N-1)}- \lambda_{i}^{(N-2)} -\frac{1}{2} )} \frac{(\lambda_{k}^{(N-1)} + \lambda_{i}^{(N-2)}  +\frac{1}{2} )}{(\lambda_{k}^{(N-1)}+ \lambda_{i}^{(N-2)} -\frac{1}{2} )} \Theta^{(N-1)} (\lambda_{k}^{(N-1)}) \nonumber \\
= \prod_{j \neq k}^{n_{N-1}} \frac{(\lambda_{k}^{(N-1)} - \lambda_{j}^{(N-1)} +1 )}{(\lambda_{k}^{(N-1)} - \lambda_{j}^{(N-1)} -1 )} \frac{(\lambda_{k}^{(N-1)} + \lambda_{j}^{(N-1)} +1 )}{(\lambda_{k}^{(N-1)} + \lambda_{j}^{(N-1)} -1 )}
\label{betre2}
\end{eqnarray}
where the function $\Theta^{(l)} (\lambda)$ is given by
\begin{equation}
\Theta^{(l)} (\lambda)= \cases{
\frac{(\lambda -\frac{l}{2} + \xi_{-})}{(\lambda +\frac{l}{2} - \xi_{-})} \frac{(\lambda -1 -\xi_{+})}{(\lambda +1 +\xi_{+})} \;\;\;\;\;\;\; l=p \cr
1 \;\;\;\;\;\;\; \mbox{otherwise} \cr}
\label{betre4}
\end{equation}

We would like to close this letter with the following remarks.  The same
strategy described above works when one of the boundaries is purely free and 
the other stays arbitrary with $2N-1$ independent parameters. This is for instance the case of 
$K_{\cal A}^{(+)}(\lambda)= \mbox{Id}$  and 
$K_{\cal A}^{(-)}(\lambda)$ an arbitrary $K$-matrix (\ref{kma},\ref{par}). It turns out that the corresponding
Bethe ansatz results for this choice of boundaries are obtained from Eqs.(\ref{einre}-\ref{betre4}) by
taking their $\xi_{+} \rightarrow \infty $ limit.  We note that similar result has recently been reported
in ref.\cite{ANAR}  however on the basis of the analytical Bethe ansatz method and solely for case of diagonal 
boundaries. 

Although we have concentrated our attention on $N \geq 3$ $SU(N)$ models, similar idea is also applicable
with success to the $SU(2)$ $\mbox{XXX}$ spin chain. In this special case it is more 
convenient to start by inserting
$M_{\cal A}^{(+)} \left [ 
M_{\cal A}^{(+)} \right ]^{-1}$   all over the double-row transfer matrix (\ref{tran},{\ref{loper}) and after
reversing the transformed $\tilde{\cal{L}}$-operators we impose  that 
$ \left [ M_{\cal A}^{(+)} \right ]^{-1} K_{\cal A}^{(-)}(\lambda) 
M_{\cal A}^{(+)}$  is a triangular matrix \footnote{This construction clearly leads us 
to fewer constrained boundary parameters than the condition $M_{\cal A}^{(+)}=M_{\cal A}^{(-)}$ but  when
$N \geq 3$ a Bethe ansatz analysis has eluded us so far.}.  This allows us to carry 
out the algebraic Bethe ansatz for a single
constraint between the six possible 
boundary parameters, reproducing what has been found earlier for the $\mbox{XXZ}$ chain \cite{CA}. However,
this method has the clear advantage of relating the eigenfunctions of the $\mbox{XXX}$ chain
having all possible boundary terms with five 
free parameters to that with only one off-diagonal and suitable
diagonal boundary terms. We hope 
that this relationship could be of utility in physical applications such as 
in the study of the scaling behaviour
of symmetric exclusion processes \cite{CHUAL}.

The possibility of undoing gauge transformed $\cal{L}$-operators appears to be a general property of
isotropic vertex models \cite{RM}. Therefore, we expect that the method devised in this work will be
useful in the solution of the eigenspectrum of a variety  of isotropic systems with certain
non-diagonal open boundaries. Interesting examples would be the case of soliton non-preserving boundaries \cite{ANAR,DOI}
for the conjugated representation of the $SU(N)$ as well as  general boundaries for vertex
models invariant by the $O(N)$ and $Sp(2N)$ Lie algebras.

\section*{Acknowledgements}
W. Galleas thanks  FAPESP (Funda\c c\~ao de Amparo \`a Pesquisa do Estado de S\~ao Paulo) 
for financial support. The work of M.J. Martins has been supported 
by the Brazilian Research Council-CNPq and FAPESP.


\addcontentsline{toc}{section}{References}
\begin{thebibliography}{}
\bibitem{SK} E.K. Sklyanin, {\em J.Phys.A:Math.Gen. 21 (1988) 2375}
\bibitem{QI}  L.A. Takhtajan and L.D. Faddeev, {\em Russian Math. Surveys, 34 (1979) 11};
V.E. Korepin, G. Izergin and N.M. Bogoliubov, {\em Quantum 
Inverse Scattering Method and Correlation Functions, Cambridge University Press, 1993}
\bibitem{RE} P.P. Kulish and N.Y. Reshetikhin, {\em  
J.Phys.A: Math.Gen. 16 (1983) L591};
O. Babelon, H.J. de Vega and C.M. Viallet, {\em Nucl.Phys.B, 200 (1982) 266}
\bibitem{DE} H.J. de Vega and A. Gonzales-Ruiz, {\em J.Phys.A:Math.Gen. 27 (1994) 6129}; 
J. Abad and M. Rios, {\em Phys.Lett.B 352 (1995) 92}
\bibitem{KU} P.P. Kulish, {\em hep-th/9507070}; M. Mintchev, E. Ragoucy and P. Sorba, {\em J.Phys.A:Math.Gen. 34
(2001) 8345}
\bibitem{NEP} L. Mezincescu and R. Nepomechie, {\em Int.J.Mod.Phys.A 6 (1991) 5231} and addendum.
\bibitem{CH} H. Fan, B.Y. Hou, K.J. Shi and Z.X. Yang, {\em Nucl.Phys.B 478 (1996) 723}
\bibitem{NEP1} R.I. Nepomechie, {\em J.Stat.Phys. 111 (2003) 1363}; {\em J.Phys.A:Math.Gen. 37 (2004) 433}
\bibitem{CA} J. Cao, H.Q. Lin, K.J. Shi and Y. Wang, {\em Nucl.Phys.B 663 (2003) 487}
\bibitem{JI} J. de Gier and P. Pyatov, {\em JSTAT 03 (2004) P002}
\bibitem{JA} W.L. Yang and R. Sasaki, {\em Nucl.Phys.B 679 (2004) 495}
\bibitem{CH1} H. Fan, B.Y. Hou, G.L. Li and K.J. Shi, {\em Phys.Lett.A 250 (1998) 79}
\bibitem{RM} G.A.P. Ribeiro and M.J. Martins, {\em nlin.SI/0406021}
\bibitem{FO} A. Foester and M. Karowski, {\em Nucl.Phys.B 396 (1993) 611}; 
H.J. de Vega and A. Gonzalez-Ruiz, {\em Nucl.Phys.B 424 (1994) 468}
\bibitem{CH2} R.H. Yue, H. Fan and B.Y. Hou, {\em Nucl.Phys.B 462 (1996) 167}; 
G.L. Li, R.H. Yue and B.Y. Hou, {\em Nucl.Phys.B 586 (2000) 711}
\bibitem{ANAR} D. Arnaudon, J. Avan, N. Cramp\'e, A. Doikou, L. Frappat and E. Ragoucy, {\em math-ph/0406021}
\bibitem{CHUAL} R.B. Stinchcombe and G.M. Sch\"utz, {\em Phys.Rev.Lett. 75 (1995) 140}; F.C. Alcaraz and
V. Rittenberg, {\em Phys.Lett.B 314 (1993) 377}.
\bibitem{DOI} A. Doikou, {\em J.Phys.A:Math.Gen. 33 (2000) 4755}


\end{thebibliography}
\end{document}